\documentclass[
 reprint,
 superscriptaddress,
 amsmath,amssymb,
prb,
]{revtex4-2}

\usepackage{physics}
\usepackage[utf8]{inputenc}
 
\usepackage{hyperref}
\usepackage{graphicx}
\usepackage{rotating}
\usepackage{amsmath,amssymb,amsfonts,textcomp}
\usepackage{color}

\usepackage{xspace}

\newcommand{\ttn}{{\tt n}}
\begin{document}

\title{Restricted Open-shell cluster Mean-Field theory for Strongly Correlated Systems }

\author{Arnab Bachhar}
\author{Nicholas J. Mayhall}
\email{nmayhall@vt.edu}
\affiliation{Department of Chemistry, Virginia Tech,
Blacksburg, VA 24060, USA}
\affiliation{Virginia Tech Center for Quantum Information Science and Engineering, Blacksburg, VA 24061, USA}

\begin{abstract}
The cluster-based Mean Field method (cMF) and it's second order perturbative correction~\cite{cmf_first}, was introduced by Jim\'enez-Hoyos and Scuseria to reduce the cost of modeling strongly correlated systems by dividing an active space up into small clusters, which are individually solved in the mean-field presence of each other. 
In that work, clusters with unpaired electrons are treated naturally, by allowing the $\alpha$ and $\beta$ orbitals to spin polarize. While that provided significant energetic stabilization, the resulting cMF wavefunction was spin-contaminated, making it difficult to use as a reference state for spin-pure post-cMF methods. 
In this work, we propose the Restricted Open-shell cMF (RO-cMF) method, extending the cMF approach to systems with open-shell clusters, while not permitting spin-polarization. While the resulting RO-cMF energies are necessarily higher in energy than the unrestricted orbital cMF, the new RO-cMF provides a simple reference state for post-cMF methods that recover the missing inter-cluster correlations. We provide a detailed explanation of the method, and report demonstrative calculations of exchange coupling constants for three systems: a di-iron complex, a di-chromium complex, and a dimerized organic radical. We also report the first perturbatively corrected RO-cMF-PT2 results as well. 
\end{abstract}

\maketitle
\section{Introduction}
Simulating open-shell systems is an important aspect of modern quantum chemistry problems due to their relevance in numerous chemical reactions, magnetic materials,~\cite{molecular_magnetism}, and electronic devices.  
Specifically, di-nuclear transition metal complexes serve as fundamental systems in molecular magnetism~\cite{molecular_magnetism}
due to the fact that their low energy spectrum consists of multiple spin states, which can mix via spin-orbit coupling, creating a barrier to spin flipping.
They further exhibit a wide range of applications spanning from catalysis~\cite{oec,nitrogenase} to materials science,~\cite{bimetallic_material} medicine,~\cite{bimetallic_medicinal} environmental protection, energy conversion,~\cite{energy_bimetallic} and sensing technologies,~\cite{nano_biosensor} etc.

Despite their importance, accurately modeling the electronic structure of transition metal systems remains a challenge. Partial occupancy in near-degenerate d-shells results in nearly degenerate electronic configurations which demands multiconfigurational treatment of the reference wavefunction. 
While Hartree-Fock (HF) provides a good starting point for most weakly correlated systems,~\cite{szabo__attila_modern_nodate,helgaker_molecular_2000-1,hartree_wave_1928}
single determinant-based traditional methods
struggle to capture most of the strong correlations in these complexes.
Density functional theory (DFT)~\cite{korth_density_2017} and truncated coupled-cluster (CC)~\cite{shavitt_many-body_2009} methods are some of the most widely used methods to capture ground- and excited-state properties.
However, both have limitations, which stem from the underlying single-determinant reference.
Moreover, DFT results can be highly functional dependent which does not allow for systematic improvements. 

As highlighted in the preceding discussion, single reference methods quickly become inaccurate for treating open-shell systems. 
As more determinants contribute significantly to the system's ground state, the HF approximation becomes a poor reference for the post-HF methods, such as truncated CC and CI. 
Because of this, multiconfigurational methods (such as complete active space self-consistent field (CASSCF)\cite{cramer2002essentials}) are the conventional approach to treating such strongly correlated open-shell systems, but factorial scaling with respect to the active space size imposes hard limits on the number of strongly correlated electrons that are able to be modeled. 


To address this factorial growth in computational cost, several methods have recently been developed that attempt to leverage locality to simplify computations.~\cite{li_block-correlated_2004, gasscf,cmf_first, parker_communication_2013, parkerCommunicationActiveSpace2014, parkerModelHamiltonianAnalysis2014, vlasscf, abraham_cluster_2021, abraham_coupled_2022, abraham_selected_2020, braunscheidel_generalization_2023} 
When molecular systems can be described in terms of inherently local chemical properties (hybridization, bond order, oxidation states, etc.), they can often be conceptualized as collections of weakly interacting moieties.
For example, in many dinuclear transition metal complexes, the two metal centers are often described by their oxidation state and local spin states. The fact that this local vocabulary can be used to interpret and predict properties of the global system (such as structure and reactivity) suggests that the various metals are relatively weakly entangled and the exact global ground state should have a relatively large overlap with a relatively small number of products of local wavefunctions (tensor product states). Similarly, systems with localized spins, such as spin-lattices where spin interactions decay with distance, also display local characteristics.
This low-entanglement structure can be revealed by directly representing many-body systems in terms of local systems, referred to here as  ``clusters", which are simply disjoint local orbital active spaces.


One can exploit these properties by representing the electronic Schr\"odinger equation in a basis of tensor product states (TPS's), where the global wavefunction is defined as a linear combination of products of locally correlated wavefunctions. 
Because TPS's already include all local (intra-cluster) dynamical correlation, the global state representation in this basis can be significantly more sparse than in the conventional (un-correlated) Slater determinant basis. 

Several notable examples of the use of tensor-product state bases are the Block-correlated coupled cluster approach of Shuhua Li~\cite{li_block-correlated_2004}, the Cluster Mean-Field theory from Scuseria and coworkers,~\cite{cmf_first,cmf_third,cMF_linear_combinations} the Active Space Decomposition method of Shiozaki and coworkers~\cite{parker_communication_2013,parker_communication_2014}, the Variational Localized Active Space Self Consistent field-State Interaction method from Gagliardi and coworkers~\cite{vlasscf,lasscf_first}, and the Tensor Product-state Selected CI and Tensor Product State-Coupled Electron Pair Approximation methods from the authors' group~\cite{abraham_selected_2020, braunscheidel_generalization_2023,abraham_coupled_2022}.

While TPS representations can indeed be effective at creating more compact wavefunctions, much of this compactness depends on the way the TPS basis is defined. 
In fact, there are many, non-equivalent,  ways to construct a TPS basis, approaches which differ in either the way the orbital clusters are defined, or the way local many-body cluster states are defined. 
For instance, one could choose to define clusters based on some localizing optimization heuristic for which there exist several options (i.e., Boys or Pipek-Mezey localization~\cite{pipek_mezey}, or a DMET-based approach~\cite{vlasscf, claudinoAutomaticPartitionOrbital2019, claudinoSimpleEfficientTruncation2019}). Furthermore, the way one chooses to define the locally correlated many-electron wavefunctions also brings about several reasonable options. For example, one could choose to use eigenvectors of the local Hamiltonian that acts on a single cluster. However, this would completely neglect interactions between clusters.  Alternatively, one could use the eigenvectors of a reduced density matrix obtained by tracing out all other clusters from an approximate global wavefunction. This is often used in tensor network state representations and also in TPSCI~\cite{abraham_selected_2020}.

Of the many ways to define a TPS basis, the cMF approach of  Jiménez-Hoyos and Scuseria\cite{cmf_first} is perhaps the most well defined, as both the orbitals and the local cluster states are uniquely defined by a single variational principle (once the sizes and occupations of the clusters are chosen by the user).
In cMF, the cluster orbitals, and cluster state coefficients are defined by variationally minimizing the energy of a single TPS wavefunction. 
The cMF method establishes a reference TPS configuration, akin to how HF serves as the reference determinant for Slater determinant-based methods. 
Also analogous to HF theory, cMF is defined by a set of stationary conditions that result in a generalized Brillouin condition:
\begin{align}
    0=& \bra{\Psi^\text{cMF}}\comm{\hat{H}}{\hat{E}^q_p}\ket{\Psi^\text{cMF}} \\
    =& \bra{\Psi^\text{cMF}}\comm{\hat{H}}{\dyad{\psi_I^{A}}{\psi_0^A}}\ket{\Psi^\text{cMF}}.
\end{align}

This approach works well for systems where all unpaired electrons are placed in the same cluster. However, when multiple clusters have ground states with non-zero net spin (e.g., multicenter organometallic complexes), defining a suitable mean field theory that preserves the spin symmetries of the global system requires more careful consideration.


The unrestricted cluster-based mean-field method (UcMF)\cite{cmf_first, papastathopoulos-katsaros_coupled_2022-1} was proposed to treat these strongly correlated spin systems, allowing each cluster to break $S^2$ but not $S_z$ symmetry.
Jiménez-Hoyos and Scuseria extended their work on cMF by introducing generalized cluster mean-field (GcMF), and $S_z$-projected generalized cluster mean-field ($S_z$GcMF).\cite{cmf_third}
GcMF allows individual clusters to break $S_z$ symmetry to get a better variational cMF energy.
$S_z$GcMF aims to restore $S_z$ symmetry while optimizing the cMF state with good symmetry quantum numbers.
The spin component that violates symmetry is eliminated, and the one that conforms to the desired symmetry is preserved by using projection operators, $\int_0^{2\pi} d\phi e^{i\phi \hat{S}_z}$.

In this paper, we propose a simple generalization of cMF to open shell systems that provides the ability to describe global states that preserve the desired spin quantum numbers $S^2$ and $S_z$. The main idea of our approach (called RO-cMF) is to generalize the cMF cost function, moving from minimizing the energy of a product of ``wavefunctions'', to minimizing the energy of a product of mixed states, where each mixed state is a statistical sum of the various spin components.
In essence, this amounts to a product of thermal states at $T=0$, where only clusters with exactly degenerate ground states lose idempotency. 
After defining the method, we apply it to a number of organometallic complexes and organic radicals.

\section{Theory}
\subsection{Tensor product space}

In most cluster-based methods, an orbital active space is partitioned into non-overlapping groups, referred to as clusters (subsets of the total available single-fermion states) based
on some desired property (e.g., locality, symmetry, etc.). We will index each of these disjoint orbital subsets, clusters, with a Roman index, $I$. 
Each cluster, $I$, supports a Fock space, for which we will define a ``cluster basis'', indexed using Greek letters, $\ket{I_\alpha}$. Generally, each cluster state will be written as a linear combination of all possible Slater determinants constructed out of the cluster's orbitals. However, this is not a strict requirement, and it is possible to use more sophisticated parameterizations of the local cluster states. 
Assuming each cluster's Fock space is un-truncated, a basis for the full \textit{global} Hilbert space can be constructed by forming all possible tensor products of local cluster states.
As previously done~\cite{abraham_selected_2020}, we further choose to define our cluster states to be eigenvectors of $\hat{N}$ and $\hat{S}_z$, which will ensure our quantum states have well-defined local quantum numbers, a feature that will simplify enforcing global symmetries.
We will then further index each cluster state with the sector of Fock space it belongs to, $\ket{I_\alpha^{\ttn_I}}$, where $\alpha$ runs over all states in the local Fock sector, ${\ttn_I}$ of the $I^{th}$ cluster.
Each global tensor product state (TPS) serving as a basis vector can be expressed using these local many-body cluster states:
\begin{align}
    \ket{\psi^{\vec{n}}}= &\ket{1_\alpha^{\ttn_1}} \ket{2_\beta^{\ttn_2}} \cdots\ket{N_\gamma^{\ttn_N}} ,
\end{align}
where $\vec{n} = \left(\ttn_1, \ttn_2, \dots, \ttn_N\right)$ is a vector index running over all possible combinations of local Fock sectors (i.e., distributions of electrons among the clusters) such that $\ket{I_\alpha^{\ttn_I}}$ spans the entire Fock space of cluster $I$.
These quantum number strings represent lists of eigenvalues of the operators \(\hat{N}\) and \(\hat{S}_z\) for each cluster in the system.\cite{abraham_cluster_2021} 
The exact wavefunction can then be expressed as a linear combination of such TPS's:\cite{abraham_selected_2020}
\begin{align}\label{eq:tps_presentation}
\ket{{\Psi}} =& 
\sum_{\vec{n}}\sum_{\alpha\in \ttn_I}\sum_{\beta\in \ttn_J}\dots\sum_{\gamma\in \ttn_N}
C^{\vec{n}}_{\alpha, \beta,\dots,\gamma}
\ket{{1_\alpha^{\ttn_\textbf{1}}}} \ket{{2_\beta^{\ttn_\textbf{2}}}} \cdots\ket{N_\gamma^{\ttn_N}},
\end{align}
where $C^{\vec{n}}_{\alpha, \beta,\dots,\gamma}$ is the coefficient tensor.
While the formalism presented is general and formally exact, as discussed in the previous section, this representation is only expected to be compact when the interactions within a cluster are stronger than those between clusters, allowing the basis vectors to incorporate a relatively large amount of electron correlation embedded within the local many-body cluster states.
As a result, the coefficient tensor, $C^{\vec{n}}_{\alpha, \beta,\dots,\gamma}$ only needs to describe inter-cluster correlation, with all intra-cluster correlation folded into the basis vectors.
The resulting wavefunction written in terms of the TPS's then requires fewer basis vectors than in the traditional Slater determinant basis.
By restricting the sum over $\vec{n}$ to only those Fock sector configurations that have the correct total number of alpha and beta electrons, we naturally preserve total particle number, $N$, and spin projection, $S_z$, symmetries in our implementation.

Following the formalism defined in the ASD method~\cite{parker_communication_2013}, the standard electronic Hamiltonian in the second quantized form:
\begin{equation}\label{eq:2nd_quantized_hamiltonian}
    \hat{H} = \sum_{pq} h_{pq}\hat{p}^\dagger \hat{q} + \frac{1}{2} \sum_{pqrs} \langle{pq|rs}\rangle  \hat{p}^\dagger \hat{q}^\dagger \hat{s} \hat{r},
\end{equation}
can be partitioned into contributions based on the number of distinct clusters involved: one-, two-, three-, and four-cluster terms.
These contributions are defined as follows:
\begin{align}\label{eq:ham}
    \hat{H} = & \sum_I \hat{H_I} + \sum_{I<J} \hat{H}_{IJ} \nonumber\\
    &+ \sum_{I<J<K} \hat{H}_{IJK} + \sum_{I<J<K<L}\hat{H}_{IJKL}.
\end{align}
In Eq. (\ref{eq:ham}), $\hat{H_I}$ includes terms where all creation and annihilation operators are within cluster $I$, $\hat{H}_{IJ}$ involves operators from both clusters $I$ and $J$, and so on.
As the ab initio Hamiltonian consists solely of two-body interactions, the maximum number of clusters involved in the interactions is limited to four.


\subsection{cluster Mean-Field Theory}

The above notation used for describing our tensor product space is general for any complete set of local cluster states. However, the compactness of global states represented in this basis ultimately is determined by the specific states chosen.
In principle, we could choose our cluster states, $\ket{I_\alpha^{\ttn_I}}$, to be those which diagonalize the cluster's local Hamiltonian, i.e., $\hat{H}_I\ket{I_\alpha^{\ttn_I}}=E_\alpha^{\ttn_I}\ket{I_\alpha^{\ttn_I}}$. 
This would then fold all local electron correlations into the cluster states.
However, the interactions between clusters ultimately affect the correlation inside of a cluster, thus neglecting all inter-cluster interactions would not create the most efficient representation. 
Adopting the formalism proposed by Jiménez-Hoyos and Scuseria\cite{cmf_first}, we choose to define our cluster states by minimizing the energy of a single TPS wavefunction:
\begin{align}
    |\psi^{\text{cMF}}\rangle=|I_0\rangle|J_0\rangle \cdots |N_0\rangle, 
\end{align}
with respect to variations in both the cluster states and the orbitals defining the clusters themselves. 
Analogous to HF theory, this provides a mean-field treatment of all inter-cluster correlations, while treating all intra-cluster correlations explicitly. 

In addition to providing the lowest energy reference TPS, this approach additionally offers a more reproducible method for defining orbital clusters rather than relying on arbitrary localization criteria as our approach is based on a variational principle.

\subsubsection{Optimization of a single TPS wavefunction}
To simplify the discussion, we first consider a concrete example of  a system consisting of two clusters, $A$ and $B$,
\begin{align}
    |\psi_0^{\text{cMF}}\rangle=\ket{A_0}\otimes\ket{B_0},
\end{align}
where $\ket{A_0}$ is the ground state of that particular cluster (note that we have suppressed the Fock sector index $\ttn_A$ for convenience). 

Analogous to  HF theory, we seek to optimize the local cluster states to minimize the energy of the global TPS, $\ket{\psi^\text{cMF}_0}$, a task which can be achieved with the following Lagrangian:
\begin{align}
    \mathcal{L}&=\langle\psi_0|\hat{H}|\psi_0\rangle-\epsilon(\langle\psi_0|\psi_0\rangle-1)\\
    &=\bra{A_0}\bra{B_0}\hat{H}\ket{B_0}\ket{A_0}
    -\epsilon\left(\ip{A_0}\ip{B_0}-1\right).\nonumber
\end{align}
Making the Lagrangian stationary with respect to linear variations in the cluster basis coefficients of cluster state $\ket{A_0}$ results in a local Schr\"odinger equation with an effective Hamiltonian, $\hat{H}^{cMF}_A$:
\begin{align}\label{eq:lagrangian_expression}
\hat{H}^{cMF}_A\ket{A_0}
    =&\epsilon\ket{A_0}, 
\end{align}

that arises from tracing out the remaining clusters: 
\begin{align}
    \hat{H}^{cMF}_A &= \Tr_B\left(\hat{H}\dyad{B_0}\right) \\
    &= \hat{H}_A + \bra{B_0}\hat{H}_{AB}\ket{B_0} + E_B\\
    &= \hat{H}_A + \mathcal{V}_{A[B]} + E_B,
\end{align}
where $\expval{\hat{H}_{AB}}{B_0} = \mathcal{V}_{A[B]}$ is the mean-field potential coming from cluster $B$ acting on cluster $A$.

While only two-body terms ($\hat{H}_{AB}$) are needed in this example with 2 clusters, for systems with multiple clusters, one might expect that three- and four-body terms would ultimately be needed. However,
because three and four-body clustered Hamiltonian terms will necessarily have an odd number of creation or annihilation operators in at least one of the clusters, their contributions will be zero.



The presence of the mean-field potential, $\mathcal{V}_{A[B]}$, results in a non-linear system of equations, where cluster states, $\ket{A_0}$ depend on $\ket{B_0}$ and vice versa. 
We can write this mean-field potential simply by considering the second quantized form of the Hamiltonian term $\hat{H}_{AB}$ as:
\begin{align}
    \mathcal{V}_{A[B]}=&\bra{B_0}\sum_{pr}^A\sum_{qs}^B\langle pq|rs\rangle \hat{p}^\dagger \hat{q}^\dagger \hat{s} \hat{r}\ket{B_0}\nonumber \\+& \bra{B_0}\sum_{ps}^A\sum_{qr}^B\langle pq|rs\rangle \hat{p}^\dagger \hat{q}^\dagger \hat{s} \hat{r}\ket{B_0}\label{eq:potential}\\
    =&\sum_{pr}^A\hat{p}^\dagger\hat{r}\sum_{qs}^B\bra{B_0}\hat{q}^\dagger\hat{s}\ket{B_0} \langle pq||rs\rangle\\
    =&\sum_{pr}^A\hat{p}^\dagger\hat{r}\sum_{qs}^BP_{qs}^B \langle pq||rs\rangle\\
    =&\sum_{pr}^A\hat{p}^\dagger\hat{r} \mathcal{V}_{pr}\label{eq:potential_last_express},
\end{align}
where $P_{qs}^B$ is the one-particle density matrix for cluster $B$.

Generalizing this to a system with an arbitrary number of clusters, the cMF Hamiltonian for cluster $I$ is expressed as :
\begin{align}\label{eq:cmf_hamiltonian}
\hat{H}^{\text{cMF}}_{I} =& 
\sum_{pq\in I}h_{pq} \hat{p}^\dagger \hat{q}+\sum_{pqrs\in {I}}\left<pq|rs\right> \hat{p}^\dagger\hat{q}^\dagger\hat{s}\hat{r} \nonumber \\
&+ \sum_{J\neq I}\sum_{pq\in {I}}\sum_{rs\in {J}}P_{rs}^J\left<pr||qs\right>\hat{p}^\dagger \hat{q},
\end{align}
where $h_{pq}$, $\left<pq|rs\right>$, $\left<pr||qs\right>$ are one-electron, two-electron, and anti-symmetrized two-electron integrals, respectively.

\subsubsection{Restricted Open-shell cluster Mean-Field Theory (RO-cMF)}
While the above works well for gapped clusters, if a single cluster has a degenerate ground state, then it becomes difficult to define the effective potential (i.e., which of the degenerate microstates should be used to compute the density matrix $P_{rs}^J$?). Degenerate ground states readily occur when a given cluster has a high-spin ground state, because all $2S+1$ microstates are degenerate. Consider, for example, a cluster, $B$, with a doublet ground state. One must decide which $M_s$ microstate should be used when computing the embedding potential $\mathcal{V}_{A[B]}$:  $\ket{S=\frac{1}{2},M_s=\{\frac{1}{2}, \frac{-1}{2}\}}$ are both options. 
If we were to use any one of these spin-polarized states, then our embedding potential would be spin-dependent, and our final cMF wavefunction will be spin-polarized. While this can be helpful for generating the variationally lowest energy TPS, we are mainly interested in using cMF as a starting point, a reference state for post-cMF calculations. As such, preserving spin symmetries is more critical to our goals than simply achieving a lower reference state energy. 
In this section, we propose an analogy to ROHF, which allows one to perform cMF calculations on open-shell systems, without generating a spin-polarized solution. 

We start by acknowledging that the fundamental problem described above arises from the requirement of choosing a single state out of a degenerate set. We could naively fix this by taking an equal superposition of all spin microstates. However, the relative phases would matter, and so one would still be stuck with the same problem. Alternatively, one could take an equal statistical mixture of the degenerate microstates. The resulting state is no longer a wavefunction, or pure state, but rather is the following mixed state:
\begin{align}\label{cluster_density}
    \rho_I&=\frac{1}{2S+1}\sum_{M_s} \rho_I^{M_s}\\
    &=\frac{1}{2S+1}\sum_{M_s}\dyad{\psi_0^{\text{FCI}};S,M_s},
\end{align}
In fact, this definition is essentially the zero kelvin thermal state, $\rho_I =\lim_{\beta\rightarrow \infty} \exp\lbrace-\beta\hat{H}_I^{cMF}\rbrace/Z$ of the associated cluster,  providing a unique, non-spin polarized, state for defining the effective potential. 
Using this we can generalize the cMF procedure, where the target state is not just an unentangled product of cluster wavefunctions, but rather  an unentangled product of zero kelvin thermal states: 
\begin{align}
    \rho^{\text{RO-cMF}}&=\rho_1\otimes\rho_2\otimes\cdots\rho_N,
\end{align}
where $\rho_I$ is the mixed state obtained by taking an equal mixture of all ground state spin microstates.
The RO-cMF energy can then be represented as:
\begin{align}
    E^{\text{RO-cMF}}=&\Tr(\rho^{\text{RO-cMF}}\hat{H})
\end{align}
While this might seem like a rather significant departure from the original cMF, the actual working equations are almost identical. The embedding potential for RO-cMF is similarly derived by simply tracing out the remaining clusters:
\begin{align}
\mathcal{V}_{I[J]}=&\Tr_J({\rho}_{J} \hat{H}_{IJ})
    =\Tr_J\left(\frac{1}{N_g}\sum_{M_s}\rho_J^{M_s}\hat{H}_{IJ}\right)\\
    =&\frac{1}{N_g}\sum_{M_s}\expval{\hat{H}_{IJ}}{J;M_s}\\
    =&\frac{1}{N_g}\sum_{M_s}\sum_{pr\in I}\hat{p}^\dagger\hat{r}\sum_{qs\in J} P_{qs}^J(M_s) \langle pq||rs\rangle\\
    =&\sum_{pr}^I\hat{p}^\dagger\hat{r}\sum_{qs}^J\tilde{P}_{qs}^J\langle pq||rs\rangle,
\end{align}
where $N_g=2S+1$, and $\ket{J;M_s}=\ket{\psi_0^{\text{FCI}};S,M_s}$.
The average one-particle density matrix for all the degenerate spin states in the ground state configuration of cluster $J$ is expressed as: 
\begin{align}
    \tilde{P}_{qs}^J=\frac{1}{N_g}\sum_{M_s}P_{qs}^J(M_s).
\end{align}

\begin{figure}[ht]
    \centering
    \includegraphics[width=1\linewidth]{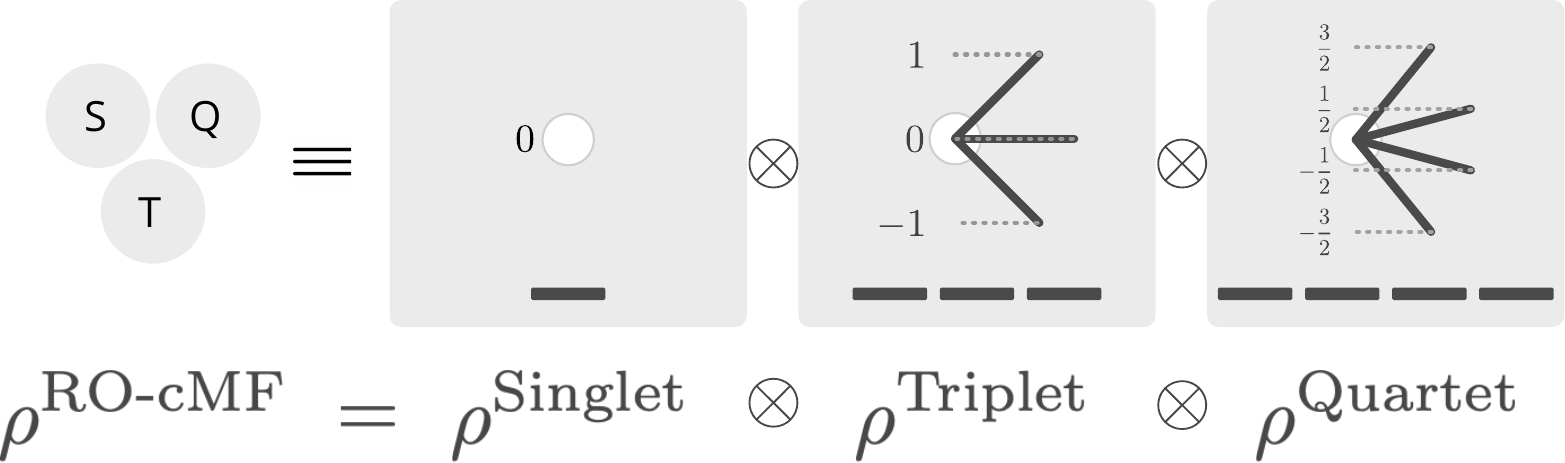}
    \caption{Pictorial depiction of the RO-cMF ground state density of a system divided into three clusters with having zero, two, and three unpaired electrons in those clusters ground states, singlet, triplet, and quartet, respectively. }
    \label{fig:ro_cmf}
\end{figure}

Thus, the RO-cMF method can be implemented in exactly the same way as non-degenerate cMF, by just replacing the ground state 1RDM, with the spin-averaged 1RDM. 
Figure \eqref{fig:ro_cmf} provides a visual representation of how the RO-cMF density is achieved for this system.
Here, the first cluster has no unpaired electron in its ground state while the other two clusters have two and three unpaired electrons, respectively.
So, the ground state density of the system will be the product of the spin densities of the mixed states of these three clusters.



\subsubsection{The RO-cMF Energy}\label{sec:rocmf_energy}
If only a single cluster has a non-zero spin, then the resulting approach is directly analogous to ROHF. 
In fact, if one created a system of 3 clusters that each had 1-dimensional Hilbert spaces: a doubly occupied cluster, a half-filled high spin cluster, and an empty cluster, then the RO-cMF optimization is equivalent to ROHF. 
However, if multiple clusters have high spin ground states, then the resulting energies are a bit more subtle. 

\begin{figure*}
    \centering
    \includegraphics[width=\linewidth]{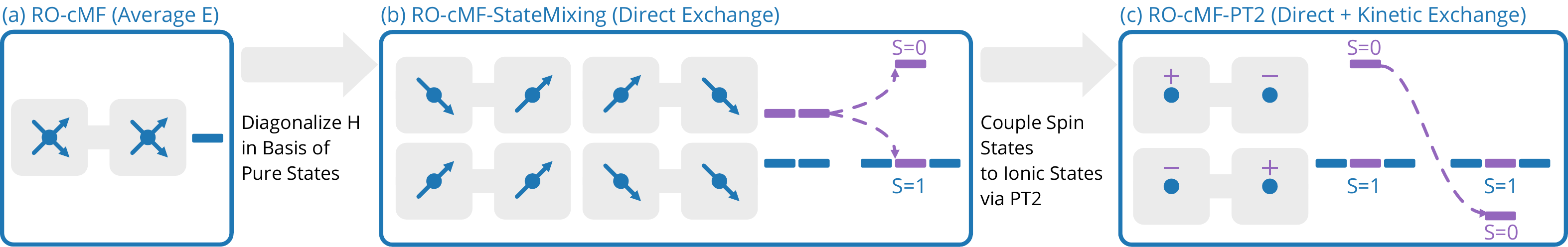}
    \caption{Schematic illustrating the way RO-cMF is used to generate low-energy spin-states. (a) RO-cMF variationally minimizes the unentangled product of mixed states obtained by averaging over all $M_s$ components of the ground states. The RO-cMF energy is the average of all the possible spin orientations. 
    (b) Diagonalizing the Hamiltonian in the basis of all spin-orientations contained in the RO-cMF mixed state. Generally direct exchange dominates, favoring ferromagnetic coupling.
    (c) Applying PT2 correction to the Spin Mixed RO-cMF states introduces new mechanisms such as kinetic exchange via coupling to ionic configurations. This generally favors antiferromagnetic coupling. }
    \label{fig:rocmf_proof}
\end{figure*}

To analyze this in a bit more detail, we can inspect a simple concrete example: a system comprised of two doublet-spin clusters, each containing 1 unpaired electron, illustrated in Fig. \ref{fig:rocmf_proof}.
While each cluster's ground state has doublet spin, the full system's ground state will re-couple these clusters into either a global singlet state or triplet state (see Fig. \eqref{fig:rocmf_proof}).
Following Eq. \ref{cluster_density}, the density of each cluster is expressed as:
\begin{align}
    \rho_A=&\rho_B=\frac{1}{2}\left(|\downarrow\rangle\langle\downarrow| + |\uparrow\rangle\langle\uparrow|\right).
\end{align}
Consequently, the RO-cMF product state is:
\begin{align}
    \rho^\text{RO-cMF} = \rho_A \otimes \rho_B = \frac{1}{4}(&\dyad{\uparrow \uparrow} + \dyad{\uparrow \downarrow} 
    \\
    &+ \dyad{\downarrow \uparrow } + \dyad{\downarrow \downarrow } ),
\end{align}
as depicted schematically in Fig. \ref{fig:rocmf_proof}(a).


If we list out the reduced density matrices for each of the global spin states, 
we see that the RO-cMF state is not a pure spin state:
\begin{align}
\rho^{0,0}=&\frac{1}{2}(|\uparrow \downarrow\rangle -|\downarrow \uparrow \rangle)(\langle\uparrow \downarrow| - \langle \downarrow \uparrow |)\\
    =&\frac{1}{2}(|\uparrow \downarrow\rangle\langle\uparrow \downarrow|+|\downarrow \uparrow \rangle\langle\downarrow \uparrow |\nonumber\\
    &-|\uparrow \downarrow\rangle\langle\downarrow \uparrow |-|\downarrow \uparrow \rangle\langle\uparrow \downarrow|),\label{eq:density_00} \\ 
\rho^{1,0}=&\frac{1}{2}(|\uparrow \downarrow\rangle\langle\uparrow \downarrow|+|\downarrow \uparrow \rangle\langle\downarrow \uparrow |\nonumber\\
    &+|\uparrow \downarrow\rangle\langle\downarrow \uparrow |+|\downarrow \uparrow \rangle\langle\uparrow \downarrow|),\\
\rho^{1,+1}=&|\uparrow \uparrow\rangle\langle\uparrow \uparrow|,\\
\rho^{1,-1}=&|\downarrow \downarrow \rangle\langle\downarrow \downarrow |\label{eq:density_1-1}
\end{align}
However, if we add all the spin microstate densities together, we generate the ``barycentric'' density matrix, 
which is easily seen to be identical to the RO-cMF state:
\begin{align}
    \rho^{\text{barycenter}} &=\frac{\rho^{0,0}+\rho^{1,+1}+\rho^{1,0}+\rho^{1,-1}}{4}\label{eq:density_avg} \\
    &= \frac{1}{4}(\dyad{\uparrow \uparrow} + \dyad{\uparrow \downarrow} 
    \\
    &+ \dyad{\downarrow \uparrow } + \dyad{\downarrow \downarrow } )\\
    &=\frac{1}{2}(|\downarrow\rangle\langle\downarrow| + |\uparrow\rangle\langle\uparrow|)\otimes \frac{1}{2}(|\downarrow\rangle\langle\downarrow| + |\uparrow\rangle\langle\uparrow|)\\
    &=\rho_A \otimes \rho_B=\rho^{\text{RO-cMF}}
\end{align}
This reveals that the energy of the RO-cMF state is equal to the barycenter of the spin-ladder generated by recoupling all open-shell clusters, 
as illustrated in Fig. \ref{fig:rocmf_proof}(b).
From the above discussion, it is evident that RO-cMF treats all the degenerate spin states on an equal footing, allowing a post-cMF treatment of inter-cluster correlations to achieve balanced descriptions of spin states, and to preserve the spin symmetry of the whole system in cluster representation.

The energy of the singlet state above is only the singlet combination of the open-shell configurations. 
Charge resonance excitations between open-shell clusters which is generally responsible for low-spin re-coupling are not included in this energy, meaning that the barycenter is typically going to be higher in energy than the high-spin state. 
After mixing all the spin states in Fig. \ref{fig:rocmf_proof}(b), one can then include these remaining interactions via perturbation theory, as depicted in \ref{fig:rocmf_proof}(c) and described in Sec. \ref{sec:pt2}, or using more accurate approaches such as TPSCI~\cite{abraham_selected_2020}. One could also use the State Interaction approach described in Ref. \cite{pandharkar_localized_2022}.

\section{Results and Discussion}\label{sec:results_discussion}

In this section, we explore the effectiveness and applicability of RO-cMF theory in treating open-shell systems with a series of systems: a transition metal di-iron complex, $[$Fe$_2$OCl$_6]^{2-}$, a dichromium complex, and organic radical, phenalenyl dimers.
The computation of exchange coupling constants in multi-center transition-metal complexes plays a key role in understanding the origins of molecular magnetism.\cite{molecular_magnetism}
The theoretical framework incorporates well-established phenomenological concepts, such as direct exchange and Anderson ligand-mediated super-exchange,\cite{superexchange1,superexchange2} which are instrumental in elucidating the observed ferromagnetic or antiferromagnetic coupling, and rationalizing experimental data.
\subsection{Heisenberg Hamiltonian}
The simplest typical model used to describe magnetic behavior is the phenomenological Heisenberg-Dirac-van Vleck (HDvV) Hamiltonian, which can be derived from the Hubbard model at half-filling using quasi-degenerate perturbation theory.\footnote{For cases away from half-filling, the system can be described by different Hamiltonians, such as the $tJ$ or double exchange models.} For systems featuring two magnetic centers, $a$ and $b$, the $\hat{H}^{HDvV}$ Hamiltonian is written simply as the product of the spin operators on two centers:
\begin{align}
    \hat{H}^{HDvV} = -2J \hat{\vec S}_a \cdot\hat{\vec S}_b,
\end{align} 
where $\hat{\vec S}_a = \hat{S}_a^x\vec x +\hat{S}_a^y\vec y +\hat{S}_a^z\vec z$, are the spin operators associated with magnetic center $a$, and $J$ is the strength of coupling between two spin centers.
The exchange coupling constant, $J$, dictates the nature of spin alignment, being positive for ferromagnetic (F) and negative for antiferromagnetic (AF) interactions, with its magnitude indicative of interaction strength.
These $J$ values that parameterize the HDvV Hamiltonian, completely determine the resulting energy spectrum of low-energy spin states, which for a two-center system is given by the Landé interval rule:
\begin{align}
    E(S) - E(S - 1) = -2SJ.
\end{align}
When derived from the Hubbard model with hopping strength $t$ and on-site coulomb repulsion $U$, the second order contribution to the exchange coupling constant is $J=\frac{-4t^2}{U}$. This so-called ``kinetic exchange'' highlights the role that electron delocalization or charge resonance (quantified by the hopping integral, $t$) has toward increasing antiferromagnetic coupling strengths. 
However, as discussed in Ref.  \citenum{faraday_discussions},  the zeroth order ab-initio Hamiltonian also contains contributions from non-local direct exchange, $K$, in addition to the second order kinetic exchange coming from the Hubbard model.
As bare direct exchange favors high-spin states and the second-order kinetic exchange term favors low-spin states, even determining the correct sign for the exchange coupling constant $J$ can be challenging also.

While $J$ is a useful quantity for rationalizing multicenter complexes, ultimately the HDvV Hamiltonian is an approximate, simplified description of the true electronic structure. 
As such, ab initio calculations are invaluable for not only computing values of $J$ from energy differences between computed states but also for determining the suitability of a phenomenological Hamiltonian for a given complex.
However, computing these low-energy states accurately is a well documented challenge for computational chemistry.
Many different factors contribute to the spin states energies~\cite{malrieuMagneticInteractionsMolecules2014} of multicenter organometallic complexes with open-shell transition metal centers.
An important factor is the modulation of magnetic interactions by the ligands.
In instances where metal ions are spatially separated by a linear bridging ligand, a “direct" metal–metal interaction is absent.
However, in cases of multiply bridged dimers, the interaction strengths can change due to the variable metal-metal distances permitted by the bridging topology.
This proximity may result in antiferromagnetic coupling strength deviating from expectations based solely on the chemical nature of the bridges, leading to intricacies in the description of the magnetic interaction.
Firstly, orbital mixing between metal and bridging ligand facilitates the delocalization of unpaired electrons and hence thereby increases the magnetic coupling.
Dynamical spin and charge polarization can further affect the coupling strengths.
Conjugated long bridging units have low-lying $\pi$-$\pi^{*}$ valence excited states, which can lead to a high degree of spin and charge polarizations.

Because of all the myriad of contributions affecting these low-energy spin states, weakly interacting metal centers with unpaired electrons exhibit high degrees of multiconfigurational complexity.
As mentioned in the introductory section, conventional approaches such as perturbation theory or coupled cluster theory are unsuitable because they rely on a qualitatively accurate single Slater determinant wavefunction as a reference.
Given that the single-determinant representation only provides direct access to the high-spin (HS) state, \(J\) values in DFT are typically derived using broken-symmetry (BS) and high-spin (HS) states~\cite{vb_antiferromagnetic,NEESE2009526}. Although this approach is routinely capable of providing qualitative accuracy, the dependency of results on the chosen DFT functionals needs careful consideration, with quantitative accuracy often depending on the specific functional employed, making it impossible to systematically improve the results.


In this article, we apply RO-cMF theory to obtain clustered representations for multi-radical systems such as these kinds of transition metal complexes, with an eye toward providing a compact reference state for post-cMF methods, such as TPSCI~\cite{abraham_selected_2020}, TPS-CEPA~\cite{abraham_coupled_2022}, etc.
We have computed the exchange coupling constants for  $[$Fe$_2$OCl$_6]^{2-}$, $[$L$_2$Cr(III)$_2(\mu-$OH$)_3]^{3+}$, $\text{L = N,N$'$,N$''$- trimethyl-1,4,7-triazacyclononane}$, and phenalenyl-dimers using both RO-cMF with State Mixing, as well as with a PT2 corrected RO-cMF-PT2. 

\subsection{Iron(III) Dimer}\label{fe_complex}

\begin{figure}
    \centering
    \includegraphics[width=1\linewidth]{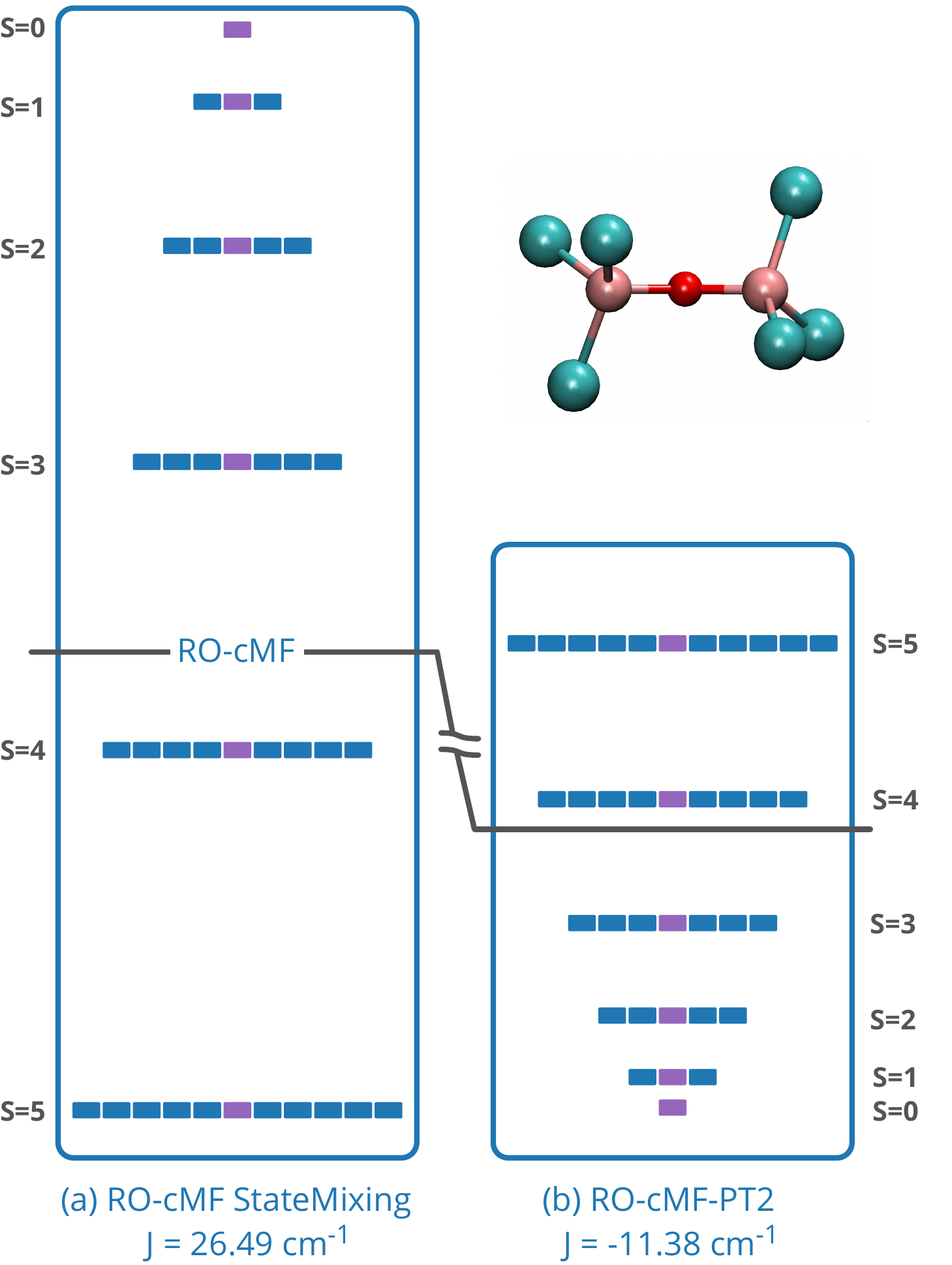}
    \caption{Spin ladder for $[$Fe$_2$OCl$_6]^{2-}$ calculated using RO-cMF and RO-cMF-PT2 relative to the undecet energies for those methods respectively. The number of dashes indicates the multiplicity of the spin state. The J value is calculated using the singlet-undecet gap in (52e, 52o) active space. 
    }
    \label{fig:spin_ladder}
\end{figure}

The exchange coupling within $[$Fe$_2$OCl$_6]^{2-}$ (see Fig. \ref{fig:spin_ladder}(a)) has been investigated with various theoretical methods, including unrestricted HF (UHF),\cite{uhf_fe} density functional theory (DFT),\cite{dft_fe1,dft_fe2}  internally contracted MRCI (IC-MRCI)~\cite{ic_mrci}, and DMRG~\cite{harris_ab_2014,block2}.
\begin{figure}
    \centering
    \includegraphics[width=1\linewidth]{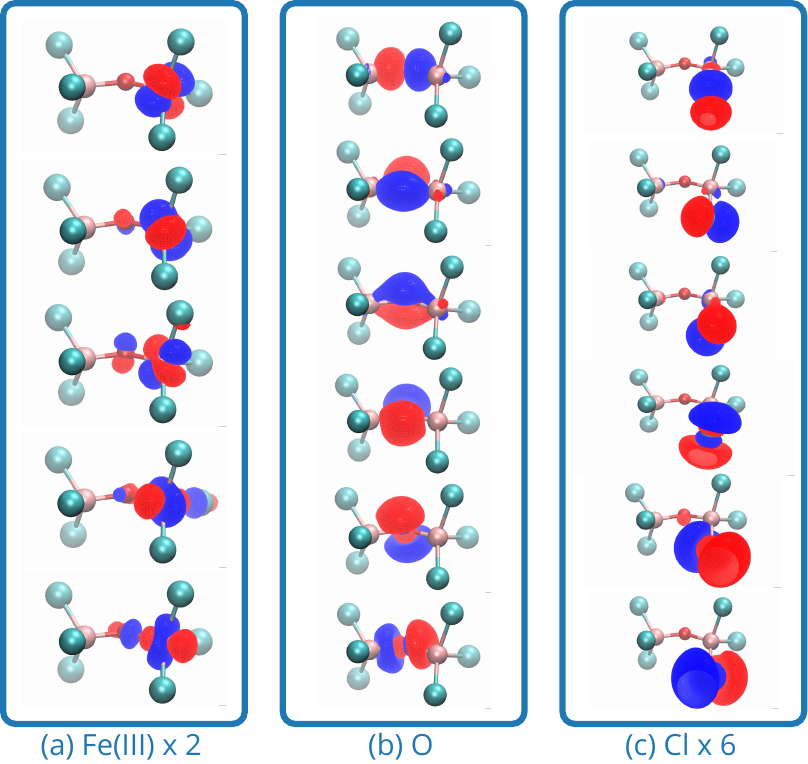}
    \caption{Molecular orbitals of active space of $[$Fe$_2$OCl$_6]^{2-}$. (a) Active space orbitals of iron atom in cluster 1 or 3, respectively. (b) active space orbitals of oxygen atom in cluster 2. (c) active space orbitals of chlorine atom in clusters 4,5, 6, 7, 8, 9. }
    \label{fig:mos_fe}
\end{figure}
In a previous study, Morokuma et al.\cite{harris_ab_2014} investigated the impact of both basis sets and active-space selection on the calculated $J$ values  for dinuclear complexes using density matrix renormalization group algorithm (DMRG),
which has become a standard benchmark method for computing exchange coupling constants in transition metal complexes 
\cite{pantazis_meeting_2019,
szalayTensorProductMethods2015,
baiardiDensityMatrixRenormalization2020,
schollwockDensitymatrixRenormalizationGroup2011,
woutersDensityMatrixRenormalization2014,
eriksenGroundStateElectronic2020,
olivares-amayaAbinitioDensityMatrix2015a,
sharmaLowenergySpectrumIronsulfur2014, 
sharmaQuasidegeneratePerturbationTheory2016a,
harris_ab_2014}.
However,  if one needs only the value of $J$ or just the highest couple spin states, spin-flip methods can be highly effective as well~\cite{mayhallComputationalQuantumChemistry2014, mayhallComputationalQuantumChemistry2015a, mayhallSpinflipNonorthogonalConfiguration2014a,pokhilkoEffectiveHamiltoniansDerived2020}.
More recently~\cite{faraday_discussions}, we have illustrated that  
TPSCI can also be used to compute low-energy states of  organometallic compounds.\cite{faraday_discussions}

One of the challenges in using active space methods is the need for the user to choose which orbitals to include. We have attempted to partially automate this process for this work. 
We start by optimizing the ROHF wavefunction for the high-spin ($S=5$) in 6-31G* basis.
We have constructed a 52 orbital active space that consists primarily of the $3d$  orbitals of each Fe center,  the $2p$ and $3p$ bridging oxygen orbitals, 
and the $3p$ and $4p$ orbitals of the six chlorine atoms.
This is done by separately projecting the occupied, open-shell, and virtual ROHF orbitals onto the associated orthogonalized atomic orbital functions, choosing the largest overlapping singular vector from each of the ROHF subspaces. 
The orbitals associated with each atomic center are grouped into separate clusters (nine in total).

Using the described clustering, we get two sextet clusters and seven singlet clusters.
As described above, to obtain the density matrices used in RO-cMF, we simply average over the 1RDMs for all degenerate $M_s$ values of each cluster's ground state. For this system, this corresponds to six $M_s$ values for the two Fe clusters ($\frac{5}{2}$, $\frac{3}{2}$, $\frac{1}{2}$, $-\frac{1}{2}$, $-\frac{3}{2}$, and $-\frac{5}{2}$) and only a single $M_s$ value for the singlet ligand clusters.
After optimizing the RO-cMF wavefunction, the resulting orbitals are shown in Fig. \eqref{fig:mos_fe}.

By diagonalizing the Hamiltonian in the relevant ($M_s=0$) sub-block (see the state mixing approach described in Fig. \ref{fig:rocmf_proof}(b)), the spin-state energies are computed and shown in Fig. \ref{fig:spin_ladder}(a).
By neglecting all ionic terms, this zeroth-order approximation incorrectly predicts a ferromagnetic alignment, with a positive $J$ value 
($J=26.5$~cm$^{-1}$) 
derived using the Landé interval rule.
When we apply perturbation theory to the RO-cMF states (Fig. \ref{fig:spin_ladder}(b), which naturally includes the inter-cluster hopping terms, the spin state ordering is reversed to the correct antiferromagnetic alignment, yielding a $J$ value of 
$-11.38$
cm$^{-1}$.%
\footnote{We note that both of these numbers correspond to the $E(S=0)-E(S=1)$ energy gap. However, we found a negligible deviation from the Land\'e intervals, and so all gaps would give nearly identical exchange coupling constants, with the largest differences on the order of $0.1$ cm$^{-1}$.}
While this is still far from the experimentally derived value of -117 cm$^{-1}$~\cite{fe2_experimental}, even the approximate inclusion of the ionic terms is able to provide qualitatively correct $J$ values. More accurate results can be obtained by treating these ionic terms non-perturbatively, as done in TPSCI~\cite{faraday_discussions}.

\subsection{Chromium(III) dimer}

Examination of chromium(III) dimers, as a paradigmatic class of antiferromagnetically coupled systems, has revealed ambiguities regarding the optimal description of their magnetic properties.
Initially, ligand-mediated superexchange\cite{cr2_superexchange} was proposed as the predominant mechanism, but empirical observations revealed a correlation between the antiferromagnetic coupling strength and the metal-metal distance in octahedral Cr$_2$(III) complexes that share faces.
This observation prompted the proposition of through-space interaction as the principal mechanism governing antiferromagnetic coupling in these complexes.\cite{noodleman_through_space,pantazis_meeting_2019}

As we mentioned in the previous section, DFT has been instrumental in analyzing exchange-coupled systems.
However, a noteworthy challenge emerges in the context of  Cr$_2$(III) complexes, where DFT reportedly falls short in providing a qualitative description of the antiferromagnetic coupling.
While BS-DFT demonstrated success in electronically similar high-valent manganese complexes, its reported failure for  Cr$_2$(III) complexes highlights the challenges in distinguishing between different coupling mechanisms, such as direct exchange versus ligand-mediated superexchange~\cite{dft_cr_failure1,case_study_cr_dimer}.
A recent study using BS-DFT, CASSCF, and DMRG by Pantazis et al.\cite{pantazis_meeting_2019} has concluded that the dominance of direct through-space interaction is corroborated, yet the additional role of superexchange introduces an additional contribution to the overall magnetic behavior.
\begin{figure}
    \centering
    \includegraphics[width=1\linewidth]{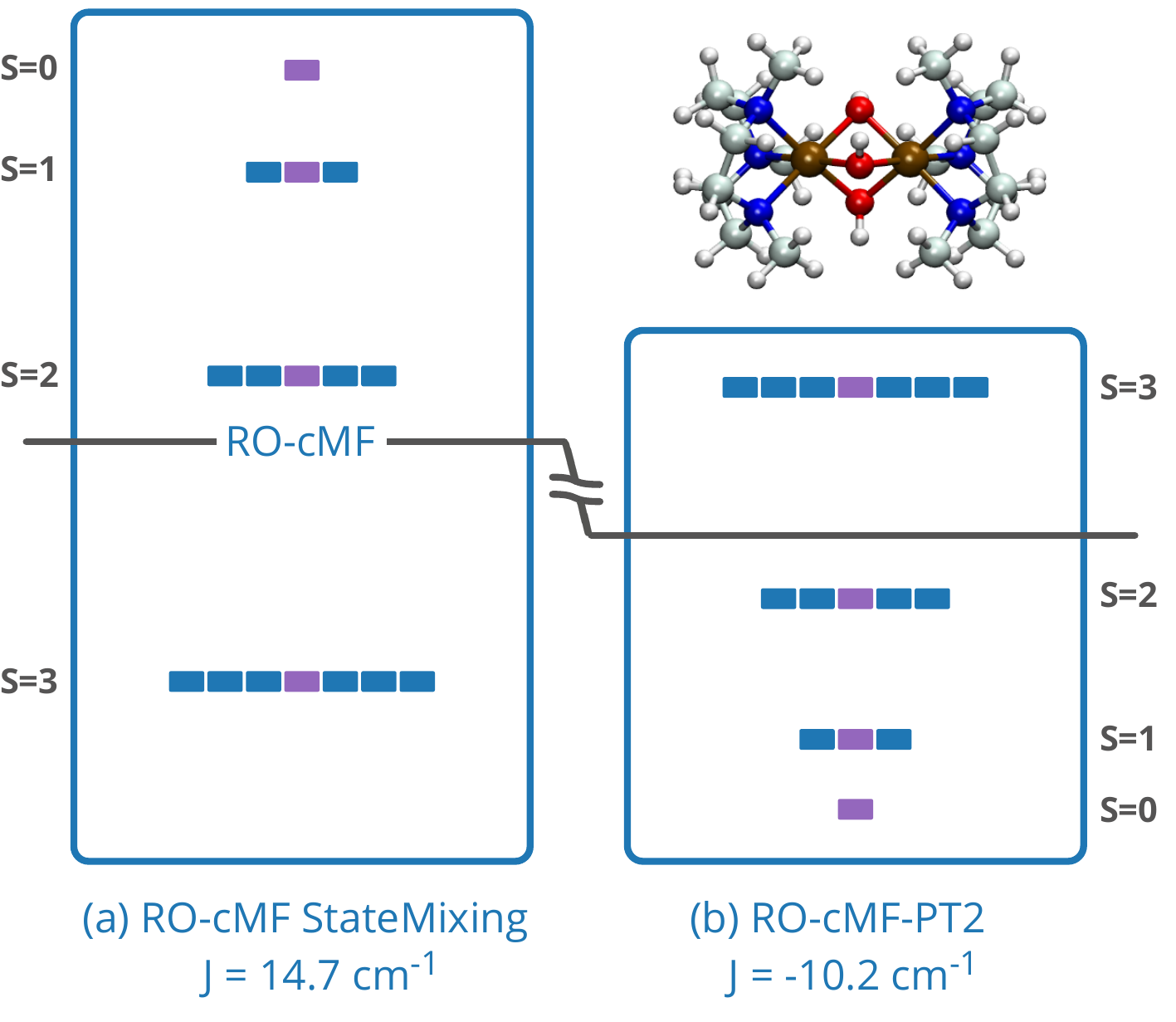}
    \caption{Spin ladder for $[$L$_2$Cr(III)$_2(\mu-$OH$)_3]^{3+}$, $\text{L = N,N$'$,N$''$- trimethyl-1,4,7-triazacyclononane}$ calculated using RO-cMF and RO-cMF-PT2 relative to the septet energies for those methods respectively. The number of dashes indicates the multiplicity of the spin state. The J value is calculated using the singlet-septet gap in (32e, 38o) active space with def2-SVP basis. }
    \label{fig:cr2_ladder}
\end{figure}

Tris-$\mu-$hydroxo Cr$_2$(III) (Fig. \ref{fig:cr2_ladder}) is a face-sharing $d^3-d^3$ complex.
For this complex, we decided to use five total clusters: one for each of the two Cr(III) centers, and one for each of the three bridging OH$^{-1}$ ligands,  yielding two quartet clusters and three singlet clusters.
Analogous to the Fe complex above,  each Cr(III) cluster is a quartet, requiring an average of the 1RDMs over all the degenerate $M_s$ microstates \{$\frac{3}{2}$, $\frac{1}{2}$, $-\frac{1}{2}$, $-\frac{3}{2}$\}. 
Using the same projection based approach, used for the Fe(III) complex, we defined each of the Cr(III) clusters by projecting onto both $3d$ and $4d$ orbitals, and each bridging OH cluster by projecting onto $2p$ and $3p$ oxygen orbitals, resulting in a total active space of (32e, 38o).
\begin{figure}
    \centering
    \includegraphics[width=1\linewidth]{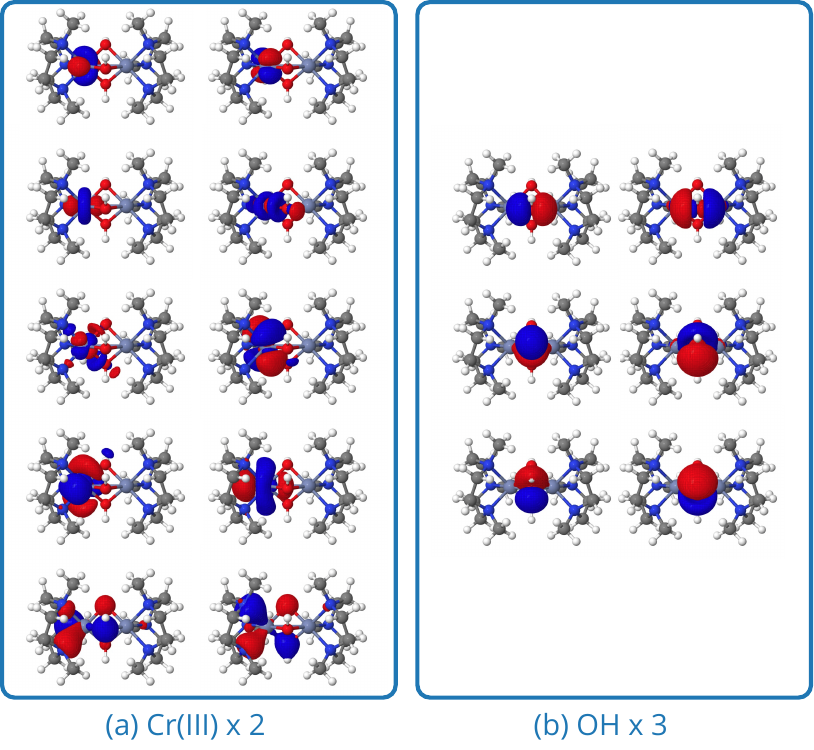}
    \caption{Molecular orbitals comprising (32e, 38o) active space for the Cr dimer complex. (a) Orbitals of (7e, 10o) Cr local active space. (b) Orbitals of (6e, 6o) bridged hydroxy group local active space.}
    \label{fig:cr2_mos}
\end{figure}
We note that during the projection procedure, the Cr(III) clusters ended up adding 2 doubly occupied orbitals, so each Cr(III) cluster is defined as a local (7e, 10o) active space instead of the expected (3e, 10o). The RO-cMF orbitals are shown in Fig. \eqref{fig:cr2_mos}.
As shown in Fig. \ref{fig:cr2_ladder}, the state mixing of RO-cMF states yields a ferromagnetic $J$ value of $14.7$ cm$^{-1}$, while RO-cMF-PT2 corrects this change this into an antiferromagnetic $J$ value of $-10.2$ cm$^{-1}$. This is nearly identical to the results obtained by Gagliardi and coworkers in Ref. ~\citenum{pandharkar_localized_2022}, using a non-perturbative vLASSCF-SI approach, on a simplified model of this same complex.
Again, while RO-cMF-PT2 is not sufficient for quantitative accuracy, the qualitative description is insightful, and can be made quantitatively accurate by more sophisticated post-cMF treatments like TPSCI, which in Ref. \citenum{faraday_discussions} found that TPSCI increased the magnitude of $J$ to $-31.3$ cm$^{-1}$, in very close agreement with CASSCF-NEVPT2 results of $-31.8$ cm$^{-1}$ \cite{pantazis_meeting_2019}.

\subsection{Organic Radicals: Phenalenyl dimer}\label{RO-cMF_radical}
Organic radicals are molecules or molecular fragments containing one or more unpaired electrons. These unpaired electrons make organic radicals highly reactive species, influencing their chemical behavior and potential applications in organic electronics, spintronics, and molecular magnetism. For example, they have been explored as components in organic light-emitting diodes (OLEDs),\cite{oled_radical} organic field-effect transistors (OFETs),\cite{ofet_radical} and organic photovoltaics (OPVs).\cite{opv_radical} 

Phenalenyl is a polyaromatic hydrocarbon $\pi$-radical characterized by its odd-alternant structure and relative stability. 
Because the unpaired electron is able to delocalize throughout the conjugated $\pi$ system, the molecule exhibits a unique stabilization relative to other organic radicals.
Despite this stability, phenalenyl radical and its substituted derivatives readily dimerize, resulting in a unique (2e, 12c) $\pi$-$\pi$ stacking bonding interaction of two radical units, termed ``pancake bonding"~\cite{kertesz_pancake_bonding}. This yields a singlet biradicaloid state where the two unpaired electrons from the multicenter SOMO orbitals of the radical monomers spin-recouple. This results in closer distances and stronger binding compared to standard dispersion bound complexes. The stabilization of the $\pi$ stacking configuration through pancake bonding renders it energetically competitive with $\sigma$-bonding. In $\pi$ stacking, the hexagonal arrangement of SOMO allows for both eclipse and staggered stacking. However, shorter $\pi$-$\pi$ bonding distance favors the staggered stacking more than eclipse which also gets destabilized due to smaller atom-atom repulsions.

\begin{figure}
    \centering
    \includegraphics[width=1\linewidth]{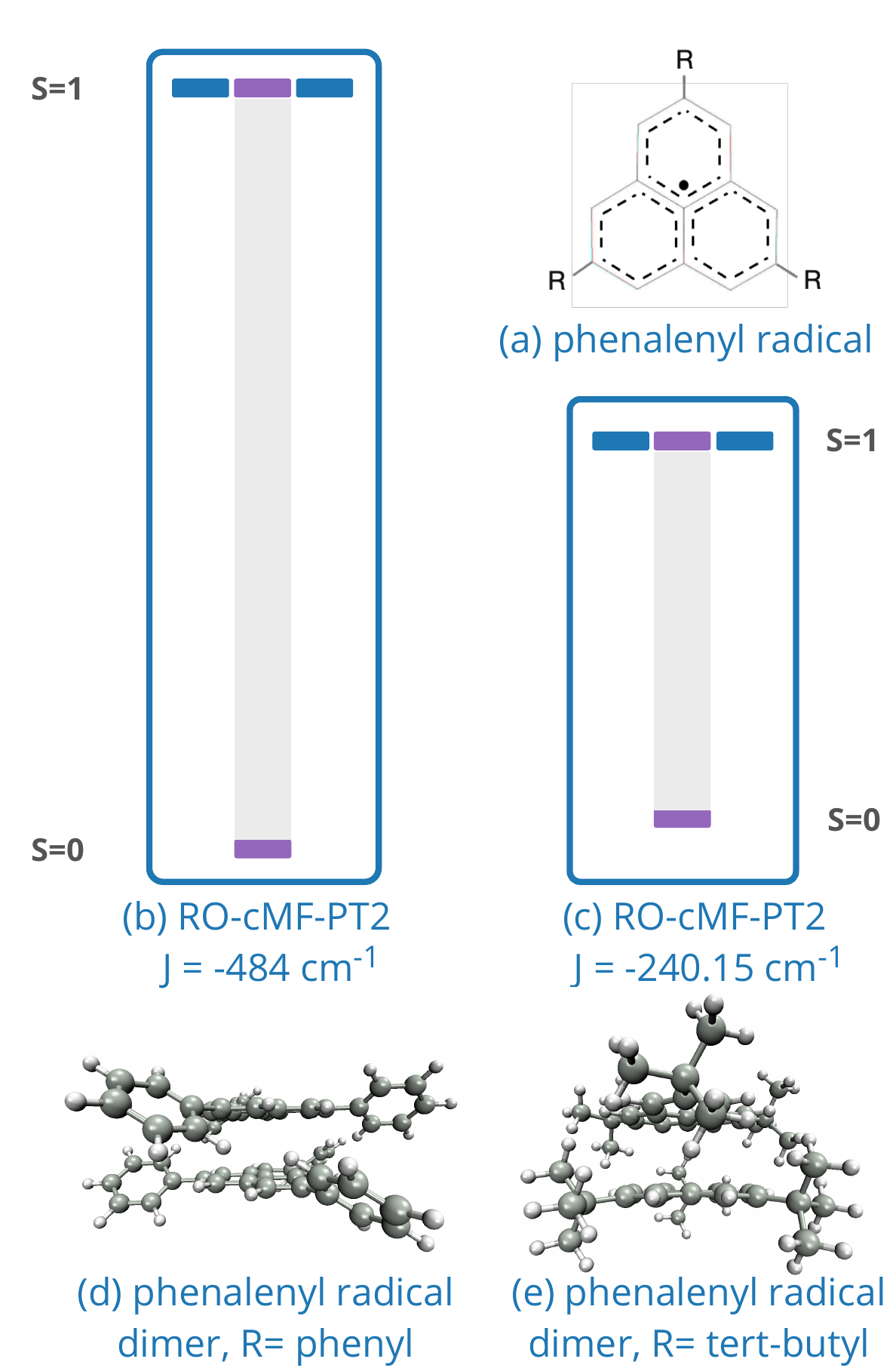}
    \caption{Relative energy levels drawn for RO-cMF-PT2 method. Exchange coupling constant, J is also shown in wavenumber. (a) Cartoon diagram of 2,5,8-R substituted phenalenyl radical. (b) Spin-state energy levels for 2,5,8-\textit{t}-butyl group substituted phenalenyl $\pi-$dimer. (c) Spin-state energy levels for 2,5,8-phenyl group substituted phenalenyl $\pi-$dimer. (d) Figure of 2,5,8-phenyl group substituted phenalenyl $\pi-$dimer. (e) Figure of 2,5,8-\textit{t}-butyl group substituted phenalenyl $\pi-$dimer.  }
    \label{fig:phenalenyl_$t$-butyl}
\end{figure}


In this section, we use RO-cMF for computing the effective exchange coupling constants for both phenyl 2,5,8-substituted and \textit{t}-butyl ($^tBu$)-substituted phenalenyl radical dimers.
In these two examples, we include all 26 $\pi$ electrons in the active space, which is then partitioned into two clusters, one for each phenalenyl radical.
For the phenyl substituted dimer, we also include the $\pi$ electrons present in phenyl R-groups, as the radical can further delocalize to some extent onto the R-groups. 
Consequently, the active space for the phenyl substituted system is (62e, 62o), broken up into six  (6e, 6o) clusters and two (13e, 13o) clusters. 
In contrast, because the unpaired electron is expected to be localized to the central phenalenyl units on the \textit{t}-butyl substituted system, we have only considered monomer radical units as clusters that make (26e, 26o) active space, partitioned into two  (13e, 13o) clusters.

Following the same procedure as with the organometallic complexes, we obtain a zeroth-order manifold of spin-states by diagonalizing the TPS's with non-zero occupations in the RO-cMF density matrix (see Fig. \ref{fig:rocmf_proof}(b)), providing spin-pure, yet physically deficient singlet and triplet states, from which an effective exchange coupling constant can be extracted.
Based on our discussion above, since the RO-cMF State Mixing Hamiltonian is constructed from only neutral states, we expected the resulting $J$ value to be positive. 
While this is true when the clusters are single Slater determinants and the only interaction is exchange, the correlation present in the local FCI states seems to provide some additional stabilization to the low-spin states, and the resulting $J$ values are correctly predicted to be negative, albeit small negative values ($-3.8$ $cm^{-1}$ and $-9.9$ $cm^{-1}$ for \textit{t}-butyl and phenyl substituted phenalenyl dimers, respectively).
Perturbative correction stabilizes the low-spin triplet state more than the high-spin singlet state which gives a $J$ value of $-240.1~\text{cm}^{-1}$ and $-484~\text{cm}^{-1}$ for \textit{t}-butyl and phenyl substituted phenalenyl dimers, respectively.
The spin-states are shown in Fig. \eqref{fig:phenalenyl_$t$-butyl}.
The perturbative correction incorporates inter-cluster hopping terms (kinetic exchange) which correctly reveals that radical units in $\pi$-dimers are strongly antiferromagnetically coupled, so much so that they form a weak covalent bond.
The difference in exchange coupling constant value arises because of the decreased distance between two carbon atoms in phenyl substituted phenalenyl dimer (3.13 angstrom) than in \textit{t}-butyl substituted one (3.5 angstrom). 
Phenyl groups help to delocalize the unpaired electron in phenyl substituted units, and this extended conjugation further seems to help the two radicals come closer to each other that increases the overlap, thus making the $J$ value more negative.
However, while the PT2 interactions are significantly stronger than the bare neutral-state interactions, comparison with TPSCI reveals that the PT2 numbers are still almost half as large as they should be 
(TPSCI yields $-403.2$ cm$^{-1}$ for \textit{t}-Bu substituted phenalenayl dimer and $-782.3$ cm$^{-1}$ for phenyl substituted complex).

\section{Conclusion}

In this paper, we have introduced the RO-cMF formalism for serving as a reference state for treating clusterable open-shell systems with tensor product state-based methods.
In order to avoid breaking spin-symmetry during the cluster-state and orbital optimization in cMF, the RO-cMF method assumes an unentangled mixed state ansatz, which is equivalent to a product of zero-Kelvin thermal states on each cluster.
The energy of this RO-cMF state then corresponds to the barycenter of the resulting spin-manifold, such that optimization minimizes all spin-states on an equal footing. 

Considering three chemical systems as examples (both a Fe(III) and a Cr(III) bimetallic compound as well as an organic radical dimer), we demonstrated how the RO-cMF method can be used as a reference state for computing low-energy spin-states in a TPS basis. By diagonalizing the Hamiltonian in the basis of TPS's that have non-zero occupations in the RO-cMF state, the resulting energy spectrum provides zeroth-order eigenfunctions that are spin-pure. Because this basis neglects many of the interactions that ultimately determine the low-energy spectrum, this zeroth-order model must be corrected by perturbation theory before qualitative accuracy can be achieved. While this approach is unable to achieve quantitative accuracy, it is intriguing as a conceptually insightful model based on the representation in terms of a natural diabatic basis.
We observe that it is generally necessary to go beyond perturbation theory (via more accurate post-cMF methods like TPSCI~\cite{abraham_selected_2020, braunscheidel_generalization_2023, abraham_coupled_2022}), if a quantitatively accurate approximation to a large active space is needed.

In future work, we plan to explore the performance of beyond-PT2 approaches like TPS-based coupled electron pair approximations for computing exchange coupling constants and excited state energies. 
All calculations used PySCF\cite{sun_recent_2020} for generating the relevant integrals, and our open-source  \href{https://github.com/nmayhall-vt/FermiCG}{FermiCG} Julia package\cite{mayhall_nicholas_fermicg_nodate} for the RO-cMF and RO-cMF-PT2 calculations.

\section{Acknowledgments} 
This work was supported by the National Science Foundation (Award No. 1752612).

\appendix

\section{Perturbation Theory}\label{sec:pt2}
Because RO-cMF does not contain any inter-cluster entanglement, the energies are rarely quantitatively meaningful. 
However, PT2 corrections can be added on top of cMF (see Ref. \citenum{cmf_first}) to provide a perturbative treatment of the missing inter-cluster correlation. 
As the RO-cMF state is formulated in terms of spin-averaged reduced density matrices, it doesn't describe a single state, but rather a manifold of low-energy spin states. 
We start by defining projectors onto the reference space, $\mathcal{P}$, (of the TPS's with non-zero occupations that have the desired $M_s$ quantum numbers) and the external space, $\mathcal{Q}=1-\mathcal{P}$.
For example, in the two-electron example described in the Section \ref{sec:rocmf_energy}, this corresponds to $\mathcal{P}=\{\ket{\uparrow \downarrow}, \ket{\downarrow \uparrow }\}$.

Next, we generate our zeroth-order reference state by diagonalizing the Hamiltonian in the basis, $\mathcal{P}$:
\begin{align}
\ket{\Psi_s^{0}} = \sum_{i\in\mathcal{P}}c^s_i\ket{\mathcal{P}_i}.
\end{align}

The total Hamiltonian, $\hat{H}$, is decomposed into a zeroth-order Hamiltonian,  $\hat{H}^{(0)}$ and a perturbed Hamiltonian, $\hat{H}^{(1)}$, using L\"owdin partitioning theory:
\begin{equation}
\begin{array}{l}
\hat{H}^{(0)}=\left(\begin{array}{c|c}
\hat{H} & 0 \\
\hline 0 & \hat{F}^{\text{cMF}}+\expval{\hat{V}^\text{cMF}}{\Psi_s^0}
\end{array}\right)
\end{array},
\end{equation}
and
\begin{equation}
\begin{array}{l}
\hat{H}^{(1)}=\left(\begin{array}{c|c}
0 & \hat{H} \\
\hline \hat{H} & \hat{V}^{\text{cMF}}-\expval{\hat{V}^\text{cMF}}{\Psi_s^0}
\end{array}\right),
\end{array}
\end{equation}
where \begin{align}
     \hat{H} &= \hat{F}^{\text{cMF}} + \hat{V}^{\text{cMF}} \nonumber \\
     \hat{F}^{\text{cMF}} &= \sum_{I} \hat{H}_I^{\text{cMF}}.\nonumber
 \end{align} 
The Fock-like cMF Hamiltonian, $\hat{F}^{\text{cMF}}$ is diagonal when excited cluster states are defined as eigenvectors of the effective local Hamiltonians, $\hat{H}^\text{RO-cMF}_I$, as is done here.
With this formulation of perturbation theory, the expression for the first-order coefficient for state $s$ takes the form:
\begin{align}
c_{j,s}^{\text{cMF(1)}} = \frac{\bra{\mathcal{Q}_j}\hat{H}\ket{\Psi_s^0}}{
\expval{\hat{F}^{\text{cMF}}}{\Psi_s^0} - \expval{\hat{F}^{\text{cMF}}}{\mathcal{Q}_j}}
\end{align}
The second order perturbation energy correction can be expressed as:
\begin{align}
    E^2_{s,\text{cMF}}=\sum_{j}c_{j,s}^{\text{cMF(1)}}\bra{\Psi_s^0}\hat{V}^{\text{cMF}} \ket{\mathcal{Q}_j}
\end{align}
While a quasi-degenerate formulation of this perturbation theory approach might become effective for the near-degenerate states~\cite{qdp_mayhall_1,qdp_mayhall_2}, this is deferred for future work.

\section{Implementation of Orbital Hessian in ClusterMeanField}

As mentioned above, the cMF wavefunction (and RO-cMF state) is optimized variationally by minimizing the energy with respect to both CI coefficients and orbital rotation parameters ($\kappa_{pq}$) in an anti-Hermitian operator, $\hat{\kappa}$.
We adopted a two-step procedure for the cMF optimization. This consists of an inner
“CI” optimization inside each cluster followed by an outer “orbital” optimization in the full system. As demonstrated in Refs. \citenum{abraham_selected_2020,cmf_first}, orbital optimization is the most significant step in minimizing the cMF energy. As shown in Ref. \citenum{abraham_cluster_2021}, the energy can be minimized by using the anti-Hermitian operator $\hat{\kappa}$, which defines the unitary transformation of the single particle basis.
\begin{align}
    \hat{\kappa}&=\sum_{p<q}\kappa_{pq}\hat{E}^{-}_{pq}\\
    &=\sum_{p<q}\kappa_{pq}(\hat{E}_{pq}-\hat{E}_{qp}),\\
    \hat{E}_{pq}&=\sum_{\sigma}a^{\dagger}_{p\sigma}a_{q\sigma},
\end{align}
where $p$ and $q$ are the orbital indices, $\sigma$ is the spin index, and the antisymmetric singlet excitation operator, $\hat{E}^{-}_{pq}$,  is expressed in terms of one-electron excitation operator, $\hat{E}_{pq}$.
The basis is rotated to a new basis upon unitary transformation, and the energy will carry orbital dependency.
\begin{align}
    &\tilde{\hat{p}}=e^{\hat{\kappa}}\hat{p}e^{-\hat{\kappa}},\\
    &E[\kappa]={\langle\psi^{\text{cMF}}|e^{\hat{\kappa}}\hat{H}e^{-\hat{\kappa}}|\psi^{\text{cMF}}\rangle}
\end{align}
We seek to optimize the cMF orbitals by finding orbital rotation parameters, $\kappa_{pq}$, that minimize $E$. 
To do this, we use a two-step Newton minimization
by expanding the energy in a Taylor series,
\begin{align}\label{eq:Taylor_energy}
    E[\kappa]=E_{\text{cMF}}+\kappa_v^{T}g_o+\frac{1}{2}\kappa_v^{T}h_{oo}\kappa_v+\cdots,
\end{align}
where $\kappa_v$ is the vectorized form of the upper triangular matrix of $\kappa_{pq}$, $g_o$ is the orbital gradient, and $h_{oo}$ is the orbital hessian.
Expanding about $\kappa_{pq}=0$, the orbital gradient $g_{o,pq}$ and orbital Hessian $h_{oo,pqrs}$ can be expressed as:
\begin{align}
    g_{o,pq}&=\frac{\delta E}{\delta \kappa_{pq}}_{\kappa=0}=\langle \psi^{\text{cMF}}|[\hat{E}^{-}_{pq},\hat{H}]|\psi^{\text{cMF}}\rangle,\\
    h_{oo,pqrs}&=\frac{\delta^{2} E}{\delta \kappa_{pq}\delta \kappa_{rs}}_{\kappa=0}\\
    &=\frac{1}{2}(1+P_{pq,rs})\langle \psi^{\text{cMF}}|[\hat{E}^{-}_{pq},[\hat{E}^{-}_{rs},\hat{H}]]|\psi^{\text{cMF}}\rangle\\
    &=G_{pqrs}-G_{qprs}=(1-P_{pq})G_{pqrs},
\end{align}
where $P_{pq,rs}$ is the permutation operator.

The final expression of the orbital gradient is
\begin{align}
    &g_{o,pq}=2(F_{pq}-F_{qp}),\\
    &F_{pq}=\sum_{\sigma}D_{p\sigma}h_{q\sigma}+\sum_{\sigma rs}d_{p\sigma rs}V_{q\sigma rs},
\end{align}
where one-particle reduced density matrix (1RDM), $D$ is contracted with the one-electron integral ($h$), and two-particle reduced density matrix (2RDM), $d$ is contracted with the two-electron integral ($V$) to compute the generalized Fock matrix, $F_{pq}$.

The $G_{pqrs}$ element in the orbital Hessian Eq. takes the form,
\begin{align}
    G_{pqrs}=\langle \psi^{\text{cMF}}|\hat{a}^{\dagger}_{p\sigma}[\hat{a}_{q\sigma},[\hat{E}^{-}_{rs},\hat{H}]]|\psi^{\text{cMF}}\rangle.
\end{align}
The $G_{pqrs}$ can be expressed in terms of a generalized Fock matrix, 1RDMs, 2RDMs, and the integrals, 
\begin{align}
    &G_{pqrs}=(1-P_{rs})(D_{pr}h_{rs}-F_{pr}\delta_{qs}+Y_{pqrs}),\\
    &Y_{pqrs}=\sum_{mn}[(d_{pmrn}+d_{pmnr})V_{qmns}+d_{prmn}V_{qsmn}],\\
    &Y_{pqrs}=Y_{rspq}
\end{align}
After combining all the equations, the final expression for the full orbital Hessian is expressed as:
\begin{align}
   h_{oo,pqrs}&=(1+P_{pq,rs}) (1-P_{pq})(1-P_{rs})\times\\
   &(D_{pr}h_{rs}-F_{pr}\delta_{qs}+Y_{pqrs})\\
   &=(1-P_{pq})(1-P_{rs})\times\\
   &[2D_{pr}h_{qs}-(F_{pr}+F_{rp})\delta_{qs}+2Y_{pqrs}].
\end{align}

\paragraph{Newton's method}
Using the cMF orbital gradient and Hessian, we can approximate the energy expression to second order, and find the optimal step by inverting the Hessian:
\begin{align}
    \kappa_{o,pq}=-h_{oo,pqrs}^{-1}.g_{o,pq}.
\end{align}
These $\kappa_{pq}$ values define a rotated orbital basis in which a new cMF wavefunction is optimized. 
This process is repeated until the norm of the orbital gradient converges to a desired tolerance.
In our implementation, a trust-radius is used to limit the max step-size of $\kappa_{pq}$, and redundant orbital rotation parameters are projected out (i.e., intra-cluster rotations).


\providecommand{\latin}[1]{#1}
\makeatletter
\providecommand{\doi}
  {\begingroup\let\do\@makeother\dospecials
  \catcode`\{=1 \catcode`\}=2 \doi@aux}
\providecommand{\doi@aux}[1]{\endgroup\texttt{#1}}
\makeatother
\providecommand*\mcitethebibliography{\thebibliography}
\csname @ifundefined\endcsname{endmcitethebibliography}
  {\let\endmcitethebibliography\endthebibliography}{}

\end{document}